\documentclass{article}

\usepackage{amssymb}
\usepackage{graphicx}
\usepackage{dcolumn}
\usepackage{bm}
\usepackage{amsmath}
\date{}
\begin{document}

\title{\bf Photon gas thermodynamics in dS and AdS momentum spaces}
\author{M. A. Gorji$^1$\thanks{{email: m.gorji@stu.umz.ac.ir}}\,,
\hspace{.2cm}V. Hosseinzadeh$^1$\thanks{{email:
v.hosseinzadeh@stu.umz.ac.ir}}\,,\hspace{.2cm} K.
Nozari$^{1}$\thanks{email:
knozari@umz.ac.ir}\,\hspace{.2cm}and\hspace{.2cm} B.
Vakili$^2$\thanks{email: b.vakili@iauctb.ac.ir (Corresponding
author)}\vspace{.15cm}\\$^1$ {\small{\it Department of Physics,
Faculty of Basic Sciences, University of Mazandaran,}}\\{\small{\it
P. O. Box 47416-95447, Babolsar, Iran}}\\$^2${\small {\it Department
of Physics, Central Tehran Branch, Islamic Azad University, Tehran,
Iran.}}}

\maketitle
\begin{abstract}
In this paper, we study thermostatistical properties of a photon gas in
the framework of two deformed special relativity models defined by the
cosmological coordinatizations of the de Sitter (dS) and anti-de Sitter
(AdS) momentum spaces. The dS model is a doubly special relativity
theory in which an ultraviolet length scale is invariant under the
deformed Lorentz transformations. For the case of AdS model, however,
the Lorentz symmetry breaks at the high energy regime. We show that
the existence of a maximal momentum in dS momentum space leads to
maximal pressure and temperature at the thermodynamical level while
maximal internal energy and entropy arise for the case of the AdS
momentum space due to the existence of a maximal kinematical energy.
These results show the thermodynamical duality of these models much
similar to their well-known kinematical duality.
\vspace{5mm}
\end{abstract}

\section{Introduction}
The first attempts in formulating quantum field theory revealed that
the divergences are the integral part of the setup. In order to
resolve this problem, Heisenberg suggested the noncommutativity
between the spacetime coordinates as $[x^\mu,x^\nu] \sim{l_{_{\rm
Pl}}^2}$, which results in a natural ultraviolet (UV) cutoff, of the
order of the Planck scale, for the system under consideration
\cite{Heisenberg}. This suggestion immediately implies a fuzzy structure
for spacetime at short distances (UV regime) such that the position
of particles cannot be determined with zero uncertainty. The
appearance of the Planck scale shows that this setup will be emerged
from the flat (non-gravitational) limit of a fundamental quantum
theory of gravity. Therefore, it seems that incorporating gravity
in quantum physics can naturally remove the divergences in quantum
field theory in one side, and also may resolve the singularities in
general relativity in the other side. The noncommutativity between
the spacetime coordinates signals a nonzero curvature for the space
of the corresponding conjugate variables that is the momentum space.
This is the Born's reciprocity conjecture which states that a
quantum theory of gravity should be formulated on curved momentum
space \cite{Born}. The first attempt in this direction was taken by
Snyder in 1947 who has formulated a noncommutative Lorentz invariant
spacetime \cite{Snyder}. The space of momenta then turns out to be
curved with de Sitter (dS) geometry and interestingly it is shown
that the quantum field theories are naturally UV-regularized in this
setup \cite{Snyder-QFT}. Recently, the relation between the seminal
work of Snyder and the noncommutativity of spacetime coordinate
became clear in the context of doubly (deformed) special relativity
(DSR) theories. Indeed, any quantum theory of gravity such as string
theory and loop quantum gravity suggest the existence of a minimal
observable length scale \cite{String,LQG}. It is therefore natural
to expect that a non-gravitational theory which supports the
existence of minimal length scale (as a natural UV cutoff for the
system) would emerge at the flat limit of quantum gravity proposal
\cite{F-QG}. In the absence of a full quantum theory of gravity one
may proceed in reverse: one starts with special relativity and
deforms it in such a way that it supports the existence of a minimal
observer-independent length scale. This is the main idea of the DSR
theories which is suggested by Amelino-Camelia \cite{DSR}. It is then
shown that there are many DSR models \cite{DSR-LI} and any model can
be understood as a different coordinatization of dS \cite{Glikman-dS1}
and also recently proposed AdS momentum spaces \cite{AdS2,AdS}. The
Snyder model then realized as a DSR model determined by a particular
basis of dS momentum space \cite{Girelli}. Evidently, the DSR models
have very different behaviors at the UV regime. For instance, some of
them predict dynamical dimensional reduction at UV regime while the
others do not \cite{DSR-DR}. In this paper, we consider two deformed
special relativity models defined by the same coordinatizations of
dS and AdS momentum spaces from the thermostatistical point of view.
The model defined on dS momentum space is indeed a DSR theory in the
sense that a UV length scale is invariant under the associated
deformed Lorentz transformations while the Lorentz symmetry breaks
at the UV regime in the AdS case. This consideration may open a new
window to compare the deformed special relativity theories such as
the DSR theories from the thermodynamical point of view.

\section{DSR theories}
In their modern formulation, DSR theories are defined on curved
momentum spaces in the context of relative locality principle
\cite{DSR-RL}. The observer-independence of the minimum length scale
implies constant curvature for the corresponding momentum space and
therefore the dS and AdS spaces are the appropriate candidates. From
the global point of view dS and AdS spaces have ${\mathbf
R}\times{\mathbf S}^3$ and ${\mathbf S}^1\times{\mathbf R}^3$
topologies respectively. Therefore, a maximal momentum and a maximal
energy will arise by the relevant identification of compact
${\mathbf S}^3$ topology with the space of momenta and ${\mathbf
S}^1$ topology with the energy space in dS and AdS momentum spaces
respectively \cite{LT}. The other identifications lead to the
non-isotropic speed of light \cite{AdS}. From the local point of
view, different coordinatizations of these curved momentum spaces
lead to the different deformed special relativity models. Depending
ons which local coordinatizations is employed, the Lorentz symmetry
may be preserved or broken at the UV regime. The DSR theories however
are defined by those coordinatizations which preserve the Lorentz
symmetry even in UV regime. Among all possible coordinatizations of
dS and AdS momentum spaces, the natural coordinate system on dS
momentum space is inspired by the bi-cross product basis of
$\kappa$-Poincar\'{e} algebra which is known as the cosmological
coordinate since it corresponds to the cosmological rendition of
dS space in position space. In this model, the Lorentz
transformations are deformed such that the Lorentz symmetry
preserves even at UV regime. Thus, this is a DSR theory. On the
other hand, its counterpart on AdS momentum space, {\it i. e.} the
model that is defined by the cosmological coordinatization of AdS
momentum space, suggests the Lorentz violation at UV regime and
therefore it is not a DSR theory. Both of these models predict
dynamical dimensional reduction from $4$ to $3$ at UV regime. We
restrict ourselves to these two models which are the same
coordinatizations of the different dS and AdS momentum spaces in
order to explore the thermodynamical properties of them.

\subsection{dS momentum space}
For the case of free particle, with which we are interested in this
paper, DSR theories on curved momentum spaces are completely defined
by the metric of momentum space and also the mass-shell (modified
dispersion relation) condition (see also Ref. \cite{DSR-AoM} for
more general case). In the case of cosmological coordinatization of
dS momentum space the metric is given by (see Ref. \cite{AdS} for
details)
\begin{equation}\label{metric-ds}
ds^2=-dE^2+\exp(2lE)\sum_{i=1}^{3}dp_i^2\,.
\end{equation}
The above line element gives the invariant integration measure\footnote{
The numerical factor $4\pi$ is considered to recover the standard
thermodynamical results at low energy regime.}
\begin{equation}\label{ds-d4p}
\frac{d\mu(E,\vec{p})}{4\pi}=\exp(3lE)dEp^2dp\,,
\end{equation}
on the momentum space. For the massless particles with which we are
interested in this paper, the deformed mass-shell relation is given
by
\begin{equation}\label{mass-shell-ds0}
{\mathcal C}\left(1-\frac{l^2{\mathcal C}}{4}\right)=0\,,
\end{equation}
in which
\begin{equation}\label{mass-shell-ds}
{\mathcal C}=-\frac{4}{l^2}\sinh^2(lE/2)+p^2\exp(lE)\,.
\end{equation}
The mass-shell condition (\ref{mass-shell-ds0}) provides two
possibilities: ${\mathcal C}=0$ and $1-\frac{l^2{\mathcal C}}{4}=0$.
The latter case leads to the dispersion relations $E_{_\pm}=-l^{-1}
\ln(-1\pm{lp})$ which gives $E_{_\pm}\approx2l^{-1}\pm{p}$ in low
energy limit $lE\rightarrow0$. Therefore, these models do not
respect the correspondence principle and we then abandon them. For
the first case with ${\mathcal C}=0$, the modified dispersion
relations are given by
\begin{equation}\label{ds-E}
E_{_\pm}=\mp{l^{-1}\ln(1-lp)}\,.
\end{equation}
In the low energy limit $lE\rightarrow0$, the above relation gives
$E_{_\pm}\approx\pm{p}$ which shows that the constraint ${\mathcal
C}=0$ is the relevant constraint for the massless particles in this
setup. The appearance of the maximal momentum $p\leq{l^{-1}}$ is
another feature of this model which is the consequence of compact
${\mathbf S}^3$ topology of the space of momenta \cite{LT}.
Moreover, equation (\ref{ds-E}) implies $E\in(-\infty,0]$ if we deal
with $E_{_-}$ and $E\in[0,\infty)$ for the case of $E_{_+}$ which
shows that the solution $E_{_+}$ is positive definite and thus is
physically relevant.

In the flat low energy limit $lE\propto{E/E_{_{\rm Pl}}}\ll1$,
the line element (\ref{metric-ds}) reduces to the flat case
$ds^2\approx-dE^2+\sum_{i=1}^{3}dp_i^2$, the invariant measure
(\ref{ds-d4p}) reduces to the standard well-known measure
$d\mu(E,\vec{p})={4\pi}dEp^2dp$, and the deformed mass-shell
condition (\ref{mass-shell-ds}) leads to the standard
Einsteinian dispersion relation ${\mathcal C}=-E^2+p^2$ with
$E,p\in[0,\infty)$.
\subsection{AdS momentum space}
In terms of physical energy and momenta $(E,p_i)$, the line element
of AdS momentum space in the cosmological coordinate is given by
\begin{equation}\label{metric-ads}
ds^2=-dE^2+\cos^2(lE)\left(\frac{dp^2}{1+l^2p^2}+p^2d\Omega^2
\right)\,,
\end{equation}
and the invariant measure on the momentum space will be
\begin{equation}\label{ads-d4p}
\frac{d\mu(E,\vec{p})}{4\pi}=\cos^3(lE)dE\frac{p^2dp}{
\sqrt{1+l^2p^2}}\,.
\end{equation}
For the massless case, the associated mass-shell condition reads
\begin{equation}\label{mass-shell-ads}
{\mathcal C}=-\frac{1}{l^2}\sin^2(lE)+p^2\cos^2(lE)=0\,,
\end{equation}
by solving of which one can easily find the modified dispersion
relations
\begin{equation}\label{ads-E}
E_{_\pm}=\pm{l^{-1}\tan^{-1}(lp)}\,.
\end{equation}
At the low energy regime $lE\rightarrow0$, we have $E_{_\pm}
\approx\pm{p}$ which shows that the setup respects the
correspondence principle. Also, $E\in(-\infty,0]$ for the case
of $E_{_-}$ and $E\in[0,\infty)$ for $E_{_+}$. We therefore again
consider $E_{_+}$ to be the appropriate solution for the
mass-shell condition. From the above relation it is clear that
there exists a maximal energy $E\leq(\pi/2)l^{-1}$ in this setup.
The relations (\ref{metric-ads}), (\ref{ads-d4p}) and
(\ref{mass-shell-ads}) reduce to their standard counterparts in
the flat low energy limit $lE\propto{E/E_{_{\rm Pl}}}\ll1$.

\section{Statistical mechanics}
In this section we are going to consider the statistical mechanics
of a photon gas in the framework of two models that are presented
in the pervious section. In Ref. \cite{DSR-Stat}, the statistical
mechanics of such theories with different coordinatizations of dS
and AdS momentum spaces is generally formulated. Here, we briefly
review the main results and then we use the setup to find the
canonical partition function for the photon gas.

In standard statistical mechanics one is dealing with an invariant
measure on a six-dimensional phase space $\Gamma_p=\Gamma_p({\vec
x};{\vec p})$ corresponding to a nonrelativistic particle. This
measure determines the number of microstates for the system by
means of which one may study the statistical mechanics in any
ensemble. The deformed special relativity theories such as the DSR
theories are, however, formulated on an eight-dimensional extended
phase space $\Gamma=\Gamma(t,{\vec x};E,{\vec p})$ for a particle
while the number of physically distinct microstates is determined
by the measure on nonrelativistic phase space. Therefore, one
should impose the mass-shell condition and also the gauge
transformation generated by it (time evolution of the system) to
obtain the invariant measure on the space of physically distinct
microstates. This can be easily deduced by using the disintegration
theorem which leads to the following invariant measure (see Ref.
\cite{DSR-Stat} for details)
\begin{eqnarray}\label{measure-p0}
\mu_p\propto\int d\mu_{p}=\int\delta(\mathcal{C})
\delta(t-t_0)d\mu(t,\vec{x})d\mu(E,\vec{p})\,,
\end{eqnarray}
where $d\mu(t,\vec{x})=dtd^3x$ is the standard Lebesgue measure on
the spacetime sector and $d\mu(E,\vec{p})=\sqrt{-g}dEd^3p$ is the
invariant measure on the momentum sector of the extended phase space
$\Gamma=\Gamma(t,{\vec x};E,{\vec p})$ with $g=g(E,{\vec p})$ being
the determinant of the metric of the momentum space \footnote{Note
that the spacetime sector has not the standard Minkowski metric,
but it is indeed the noncommutative $\kappa$-Minkowski spacetime
($\kappa\sim{l^{-1}}$ in our notation) \cite{DSR-NC}. Defining an
appropriate measure which respects all the desired symmetries on
this noncommutative spacetime is not an easy task (see for instance
Ref. \cite{Agostini} where it is shown that the standard Lebesgue
measure leads to the noncyclic action). The standard Lebesgue
measure however respects the $\kappa$-Poincar\'{e} symmetries and
also the correspondence principle which is necessary for our aim in
this paper. Therefore, we work with the Lebesgue measure on the
spacetime sector of the extended phase space $\Gamma$.}. Although
the measure (\ref{measure-p0}) restrict the eight-dimensional
extended phase space $\Gamma$ to a six-dimensional
nonrelativistic phase space, it is indeed not uniquely defined. For
instance, one could also consider $\delta({\mathcal C}^2)$ instead
of $\delta({\mathcal C})$ which is again consistent but leads to the
different statistical results! At the first glance, one could see
that substituting even the standard Einsteinian dispersion relation
${\mathcal C}=-E^2+p^2$ in (\ref{measure-p0}), the delta function
decomposes to two separate branches and one of them is indeed
corresponding to the negative energies which are irrelevant. More
generally, $\delta({\mathcal C})$ in (\ref{measure-p0}) decomposes
as $\delta({\mathcal C})=\frac{\delta(E-E_{_{+}})}{|\frac{
d{\mathcal C}}{dE}|_{E=E_{_{+}}}}+\frac{\delta(E-E_{_{-}})}{
|\frac{d{\mathcal C}}{dE}|_{E=E_{_{-}}}}$. Note that only the
positive definite energies determined by the solution $E_{_+}$ are
physically relevant. In order to restrict ourselves to the positive
energies, we thus consider the step function $\theta(E)$. The other
important issue is the correspondence principle according to which
we should recover the well-known results of standard statistical
mechanics at the low energy (or temperature) regime. In order to do
so, we should replace $\delta({\mathcal C})$ with ${|\frac{
d{\mathcal C}}{dE}|}_{{\mathcal C}=0}\,\delta({\mathcal C})$. The
correct measure which only includes the positive energies and
respects the correspondence principle, then uniquely determined and
is given by
\begin{eqnarray}\label{measure-p}
\mu_p=\int d\mu_{p}=\int\theta(E)\Big{|}\frac{d{\mathcal C}
}{dE}\Big{|}_{{\mathcal C}=0}\delta(\mathcal{C})\delta(t-t_0)
d\mu(t,\vec{x})d\mu(E,\vec{p})\,.
\end{eqnarray}
Having the measure (\ref{measure-p}) at hand, we are adequately
equipped to study the statistical mechanics for deformed special
relativity theories in any ensemble. In canonical ensemble, the
system is supposed to be in thermal bath with temperature $T$ and
the ensemble density is given by the Boltzmann factor. The
associated single-particle partition function is then given by
\begin{eqnarray}\label{pf-def}
\mathcal{Z}_1=\frac{1}{h^3}\int{d}\mu_{p}\exp(-E/T)\,,
\end{eqnarray}
where $d\mu_p$ is defined in (\ref{measure-p}). The total partition
function $\mathcal{Z}_N$ of a $N$-particles system can be written as
$\mathcal{Z}_N=\mathcal{Z}_1^N/N!$, when particles are assumed to be
kinematically and dynamically decoupled. All the thermodynamical
quantities then can be derived from $\mathcal{Z}_N$ by the standard
definitions.

Before obtaining the deformed partition functions for the photon gas
in deformed special relativity models with dS and AdS momentum
spaces, it is useful to obtain the well-known partition function of
the photon gas in standard special relativity within the constructed
setup. In the framework of standard special relativity, the measures
are the standard Lebesgue measures $d\mu(t,{\vec x})=dtd^3x$ and
$d\mu(E,{\vec p})=dEd^3p$, and the mass-shell condition for the
massless particles is given by the standard Einsteinian dispersion
relation $-E^2+p^2=0$ which has the solutions $E_{_\pm}=\pm{p}$.
Substituting in relation (\ref{pf-def}), the partition function for
the photon gas will be
\begin{align}\label{pf-sr}
\mathcal{Z}^{^{SR}}_1&=\frac{1}{h^3}\int d\mu_p\exp(-E/T)=\frac{
1}{h^3}\int{dt}dEd^3xd^3p\,\theta(E)\Big{|}\frac{d{\mathcal C}}{
dE}\Big{|}_{{\mathcal C}=0}\delta({\mathcal C})\delta(t-t_0)
\exp(-E/T)\nonumber\\&=\frac{4\pi{V}}{h^3}\int\int{dEp^2dp}\,
\theta(E)|2E|_{{\mathcal C}=0}\left[\frac{\delta(E-p)}{|2E|_{_{
E=p}}}+\frac{\delta(E+p)}{|2E|_{_{E=-p}}}\right]\exp(-E/T)
\nonumber\\&=\frac{4\pi{V}}{h^3}\int\int{dEp^2dp}\,\delta(E-p)
\exp(-E/T)=\frac{4\pi{V}}{h^3}\int_{0}^{\infty}E^2dE\exp(-E/T)=
\frac{8\pi{V}T^3}{h^3}\,,
\end{align}
which is nothing other than the result in the standard statistical
mechanics which leads to the usual thermodynamics of a photon gas.
The nontrivial effects may occur when the measure $d\mu_{X}$ or/and
the constraint $\mathcal{C}$ is deformed. In theories with minimal
length that formulated on the reduced (non-relativistic) phase space
such as generalized uncertainty principle \cite{GUP-THR},
noncommutative reduced phase spaces \cite{NC-THR} and polymerized
phase spaces \cite{Poly-THR}, always one of the phase space measure
or the dispersion relation is modified (it depends on what one
prefers to work, in canonical (Darboux) or noncanonical charts on
the reduced phase space \cite{LT,Poly-IG,Polymer-HS}). In deformed
special relativity theories such as the DSR models, depending on
the coordinate that one implements, both the measure and the
dispersion relation can be simultaneously modified
\cite{Glikman,DSR-THR}.

Using the deformed Hamiltonian constraint (\ref{mass-shell-ds}) and
substituting the invariant measure (\ref{ds-d4p}) into the
definition (\ref{pf-def}), the canonical partition function for the
photon gas in cosmological coordinatization of dS momentum space
takes the form
\begin{align}\label{pf-ds}
{\mathcal Z}^{^{DSR-dS}}_1&=\frac{4\pi}{h^3}
\int\int{dt}d^3x\delta(t-t_0)\int\int{dE}\exp(3lE)p^2dp
\times\theta(E)\Big{|}\frac{d{\mathcal C}}{dE}\Big{|}_{
{\mathcal C}=0}\delta({\mathcal C})\exp(-E/T)\nonumber\\
&=\frac{16\pi{V}}{l^2h^3}\int_0^{\infty}{dE}\sinh^2(lE/2)
\exp\Big(-\frac{(1-2lT)E}{T}\Big)=\frac{8\pi{V}T^3}{h^3
}(1-6lT+11l^2T^2-6l^3T^3)^{-1},
\end{align}
where we have substituted ${\mathcal C}$ from (\ref{mass-shell-ds}).
The integral in the above relation is evaluated over the allowed
domain $T^{-1}-2l>0$ which shows that there is a maximal temperature
\begin{equation}\label{max-temperature}
T<T_{\max}=\frac{1}{2l}\propto{T_{_{\rm Pl}}}\,,
\end{equation}
for the photon gas in this setup. This result (with a different
approach) is also obtained in Ref. \cite{Glikman} where the first
time the statistical mechanics for DSR theories was studied. In the
same manner, using the constraint (\ref{mass-shell-ads}) and
substituting the associated invariant measure (\ref{ads-d4p}) into
the definition (\ref{pf-def}), the canonical partition function for
the photon gas in cosmological coordinatization of AdS momentum
space turns out to be
\begin{align}\label{pf-ads}
{\mathcal Z}^{^{DSR-AdS}}_1&=\frac{4\pi}{h^3}\int\int{
dt}d^3x\delta(t-t_0)\int\int{dE}\frac{\cos^3(lE)}{
\sqrt{1+l^2p^2}}p^2dp\times\theta(E)\Big{|}\frac{
d{\mathcal C}}{dE}\Big{|}_{{\mathcal C}=0}
\delta({\mathcal C})\exp(-E/T)\nonumber\\&=\frac{\pi{
V}}{l^2h^3}\int_0^{\frac{\pi}{2l}}dE\sin^2(2lE)
\exp(-E/T)=\frac{8\pi{V}T^3}{h^3}\bigg(\frac{1-
\exp[-{\pi/2lT}]}{1+16l^2T^2}\bigg)\,,
\end{align}
in which we have used the modified dispersion relation
(\ref{mass-shell-ads}). The deformed partition functions
(\ref{pf-ds}) and (\ref{pf-ads}) can be rewritten in a compact
form as
\begin{align}\label{pf}
{\mathcal Z}^{^{DSR}}_1(l;V,T)={\mathcal Z}^{^{SR}}_1(V,T)\,f(lT)\,,
\end{align}
where ${\mathcal Z}^{^{SR}}_1(V,T)=\frac{8\pi{V}T^3}{h^3}$ is the
partition function of the photon gas in the standard special
relativity (see (\ref{pf-sr})) and
\begin{eqnarray}\label{f}
f(lT)=\left\{
\begin{array}{ll}
(1-6lT+11l^2T^2-6l^3T^3)^{-1}, & \hspace{0.3cm}\mbox{dS}\\
(1-\exp(-\pi/2lT))(1+16l^2T^2)^{-1}. & \hspace{0.2cm}\mbox{AdS}
\end{array}
\right.
\end{eqnarray}
It is seen that all quantum gravity effects are summarized in the
function (\ref{f}). In the low temperature limit
$lT\propto{T/T_{_{\rm Pl}}}\ll1$, where these effects are negligible
the function (\ref{f}) tends to unity, $\lim_{lT\rightarrow0}f(lT)=
1$, and relation (\ref{pf}) gives ${\mathcal Z}^{^{DSR}}_1(l;V,T)=
{\mathcal Z}^{^{SR}}_1(V,T)$, which shows that the result of
standard special relativity is recovered in this limit.

In the absence of quantum correlations, the total partition function
for the $N$-particles system in Maxwell-Boltzmann statistics will be
\begin{equation}\label{pf-tot}
{\mathcal Z}^{^{DSR}}_N=\frac{1}{N!}\left({\mathcal Z}^{^{SR}}_1\right)^N
f^N={\mathcal Z}^{^{SR}}_Nf^N,
\end{equation}
where ${\mathcal Z}^{^{SR}}_N=\big({\mathcal Z}^{^{SR}}_1\big)^N/N!$
is the total partition function in the standard special relativity.

\section{Thermodynamics}
\begin{figure}
\flushleft\leftskip+3em{\includegraphics[width=3in]{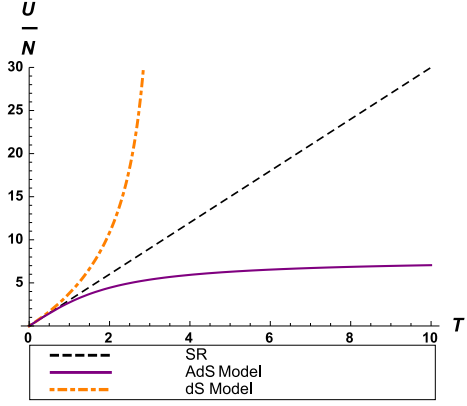}}
\hspace{3cm}\caption{\label{fig:1} Internal energy versus
temperature. For dS momentum space, the internal energy diverges at
$T_{max}$ given by (\ref{max-temperature}). The internal energy in
the case of AdS momentum space, however, gets its maximum value when
the temperature goes to infinity. These results also show the
thermodynamical duality between these two relevant models of the
deformed special relativity models. At the low temperature regimes,
both of them reduce to the standard result of special relativity.
The figure is plotted for $h=k_{_B}=c=1$.}
\end{figure}

\begin{figure}
\flushleft\leftskip+3em{\includegraphics[width=3in]{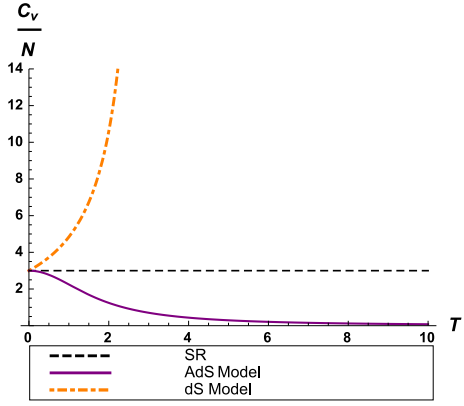}}
\hspace{3cm}\caption{\label{fig:2} The figure shows the variation
of the specific heat in terms of temperature. While this is a
temperature-independent quantity in special relativity, it
significantly changes with temperature at the high temperature
regime in deformed special relativity framework. For the dS
momentum space, it diverges at the maximal temperature
(\ref{max-temperature}) and it tends to zero when the temperature
goes to infinity in the case of AdS momentum space.}
\end{figure}

From the total partition function (\ref{pf-tot}), one can derive all the
thermodynamical quantities. The internal energy $U=T^2\frac{\partial{\ln
{\mathcal Z}^{^{DSR}}_N}}{\partial{T}}$ works out to be
\begin{eqnarray}\label{u}
U^{^{DSR}}=U^{^{SR}}+NT^2(\ln{f})'\,,
\end{eqnarray}
where a prime denotes derivative with respect to the temperature. The
first term in the right hand side, $U^{^{SR}}=3NT$, is nothing but the
internal energy of the photon gas in standard special relativity and
the second term comes from the quantum gravity corrections. The
specific heat $C_{v}=\Big(\frac{\partial U}{\partial T}\Big)_{V}$ in
this setup is obtained as
\begin{align}\label{cv}
C_{v}^{^{DSR}}=C_{v}^{^{SR}}+2NT(\ln{f})'+NT^2(\ln{f})''\,,
\end{align}
where again $C_{v}^{^{SR}}=3N$ is the well-known specific heat in
the standard special relativity and the two other terms in the right
hand side of the above relation show the quantum gravity effects. In
figures \ref{fig:1} and \ref{fig:2} we have plotted the internal
energy and the specific heat versus temperature for both the dS and
AdS momentum spaces. A comparison with their standard counterparts
in special relativity is also shown in these figures.
The quantum gravity effects will become important just at high
temperature regime. However, one can obtain a bound on the quantum
gravity scale $l$ or estimate the magnitude of these corrections
at the accessible temperatures \cite{QGExperiment}. The quantum
gravity corrections to the heat capacity of the photon gas in this
setup are
\begin{equation}\label{cv-estimation}
\frac{{\triangle}C_{v}}{C_{v}^{^{SR}}}=\frac{C_{v}^{^{DSR}}-
C_{v}^{^{SR}}}{C_{v}^{^{SR}}}=\frac{1}{3}\left(T^2(\ln{f})'
\right)'\sim{\pm{lT}}\,,
\end{equation}
which are too small to be detected by the accessible energy
scales.

Using the standard definition $P=T\frac{\partial\ln{\mathcal Z}_N
}{\partial V}$, the equation of state can be obtained as
\begin{equation}\label{eos}
PV=NT\,,
\end{equation}
which shows that the form of equation of state preserves in deformed
special relativity framework. In the Maxwell-Boltzmann statistics,
the form of equation of state in all phenomenological approaches to
the minimal length is also preserved and it seems that this is a
general feature (see for instance Refs.
\cite{GUP-THR,NC-THR,Poly-THR,Poly-IG}). An interesting result in
this setup is that there is a maximal pressure for the photon gas
for the case of dS momentum space. Existence of a maximal temperature
given by the relation (\ref{max-temperature}) immediately implies a
maximal pressure as
\begin{equation}\label{max-pressure}
P_{\max}=\left(\frac{N}{V}\right)\,T_{\max}\,,
\end{equation}
for the photon gas (note that while $N$ and $V$ are extensive
quantities, $P$ and $T$ are intensive). As far as we know,
among the other phenomenological approaches to the issue of minimal
length scale, existence of a maximal temperature and consequently an
upper bound (\ref{max-pressure}) on the pressure are only suggested
by the deformed special relativity models which are defined on the
dS momentum space.

\begin{figure}
\flushleft\leftskip+3em{\includegraphics[width=3in]{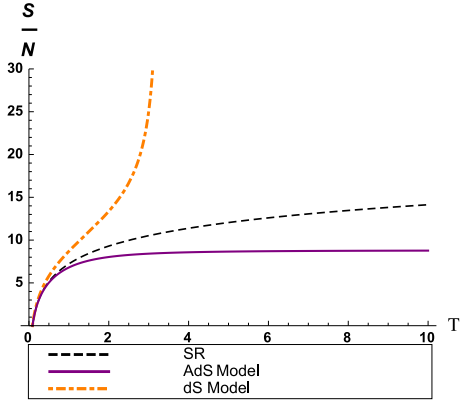}}
\hspace{3cm}\caption{\label{fig:3} The figure shows the behavior of
entropy versus temperature. As is clear from the figure, in the case
of dS momentum space, the entropy increases with increasing rate and
finally diverges at the maximal temperature. It increases forever
but with decreasing rate in the standard special relativity. In the
case of AdS momentum space, it also increases but with a decreasing
rate lower than the case of standard relativity and it has
asymptote when the temperature goes to infinity.}
\end{figure}

The entropy can be also obtained from its standard definition
$S=U/T+ \ln{\mathcal Z}_N$ which gives
\begin{align}\label{s}
S^{^{DSR}}=S^{^{SR}}+N\ln{f}+NT(\ln{f})'\,,
\end{align}
in which $S^{^{SR}}$ denotes the standard entropy of the photon gas
in special relativity and the other terms are modifications due to
the quantum effects of gravity. In figure \ref{fig:3} we have
plotted the entropy versus temperature. As this figure shows, the
entropy, in standard special relativity, increases with decreasing
rate when the temperature increases. However, it increases with
increasing rate in the case of DSR with dS momentum space and
finally it diverges when the temperature approaches to the maximal
temperature (\ref{max-temperature}). For the case of the AdS
momentum space, it increases with decreasing rate lower than the
standard one at the high temperature regime and finally it
approaches to a maximal value when the temperature goes to infinity.

All the thermodynamical features of the photon gas in the two
deformed special relativity models with dS and AdS momentum spaces
are summarized in Table~\ref{tab:1} and they are compared with each
other and also with the standard special relativity results.

\begin{table*}
\caption{\label{tab:1} Thermodynamical results of photon gas in the
deformed special relativity models defined by cosmological
coordinatization on dS and AdS momentum spaces are presented and
they are compared with each other and also with the standard
results of the special relativity:}\vspace{0.2cm}
\begin{tabular}{ccccccccc}
&Topology of&Maximal&Maximal&Lorentz&Maximal&Maximal&Maximal&Maximal\\
&Momentum&Momentum&Energy&Invariance&Internal&Entropy&Pressure&
Temperature\\&Space&&&&Energy&&&\\
\hline SR&${\mathbf R}^4$&No&No&Yes&No&No&No&No\\dS Model&${\mathbf R}\times
{\mathbf S}^3$&Yes&No&Yes&No&No&Yes&Yes\\AdS Model&${\mathbf S}^1\times
{\mathbf R}^3$&No&Yes&No&Yes&Yes&No&No\\
\end{tabular}
\end{table*}

\section{Summary and conclusions}
The DSR theories are the most well-known candidates for the flat
limit of quantum gravity proposal. There is an observer-independent
length scale (preferably of the order of the Planck length) in these
setups which leads to a natural UV cutoff for the system under
consideration. These theories are generally formulated on curved dS
and AdS momentum spaces with ${\mathbf R}\times{\mathbf S}^3$ and
${\mathbf S}^1\times{\mathbf R}^3$ topologies respectively. The
various DSR theories then can be realized from the different
coordinatizations of these curved momentum spaces. At the
kinematical level, a maximal momentum and maximal energy arise in
dS and AdS momentum spaces respectively through the relevant
identification of compact ${\mathbf S}^3$ topology with the space
of momenta in dS space and compact ${\mathbf S}^1$ topology with
the energy space in AdS case. In this respect, these two spaces are
kinematically dual to each other. In this paper, we have studied the
thermodynamical properties of a photon gas in the framework of two
different deformed special relativity models defined by cosmological
coordinatizations of dS and AdS momentum spaces. The model defined
on the dS momentum space is a DSR theory in the sense that it
preserves the Lorentz symmetry even at the UV regime. The AdS model
is however a deformed special relativity model which supports the
existence of a UV length scale but cannot preserve the Lorentz
symmetry at the UV regime. The results show that the thermodynamical
properties of the photon gas are significantly modified at the high
temperature regime such that all thermodynamical quantities enhance
in the case of dS momentum space and saturate for the AdS case. We
found that the existence of maximal momentum in dS momentum space
leads to the maximal pressure and maximal temperature at the
thermodynamical level while maximal internal energy and maximal
entropy emerged in the AdS momentum space due to the existence of
maximal kinematical energy. In this respect these spaces are
thermodynamically dual to each other much similar to their
well-known kinematical duality. All of these kinematical and
thermodynamical UV cutoffs are originated from the compact topologies
${\mathbf S}^3$ and ${\mathbf S}^1$ associated to the dS and AdS
momentum spaces. Therefore, although, we have considered a particular
case of photon gas in canonical ensemble, it seems that these results
are common feature of dS and AdS momentum spaces. Moreover, there are
many deformed special relativity models defined by different
coordinatizations of dS and AdS momentum spaces and there is not
a clear reason to prefer one from the other. Our consideration,
however, opens a new window to compare these models from the
thermodynamical point of view.

\end{document}